\documentclass[twocolumn]{aastex631} %%linenumbers,

\newcommand{\tmop}{\rm}
\usepackage{CJK}
\usepackage{threeparttable}
\usepackage{amsmath}
\makeatletter
\def\acknowledgements{%
  \par\vskip6pt\noindent%
  \section*{Acknowledgments}%
  \global\let\linenumbers\relax  % 关键：局部禁用 linenumbers
  \ignorespaces
}
\makeatother

\begin{document}
\begin{CJK*}{UTF8}{gbsn}
\title{The Dust Echo Emission of Fast Blue Optical Transients and Application to the Near-Infrared Excess of AT 2018cow}

\author[0009-0004-9719-272X]{Jing-Yao Li(李京谣)}
\affiliation{Institute of Astrophysics, Central China Normal University, Wuhan 430079, China; \url{yuyw@ccnu.edu.cn}}
\affiliation{Education Research and Application Center, National Astronomical Data Center, Wuhan 430079, China}
\affiliation{Key Laboratory of Quark and Lepton Physics (Central China Normal University), Ministry of Education, Wuhan 430079, China}

\author[0000-0002-1067-1911]{Yun-Wei Yu(俞云伟)}
\affiliation{Institute of Astrophysics, Central China Normal University, Wuhan 430079, China; \url{yuyw@ccnu.edu.cn}}
\affiliation{Education Research and Application Center, National Astronomical Data Center, Wuhan 430079, China}
\affiliation{Key Laboratory of Quark and Lepton Physics (Central China Normal University), Ministry of Education, Wuhan 430079, China}

\author[0000-0002-8708-0597]{Liang-Duan Liu(刘良端)}
\affiliation{Institute of Astrophysics, Central China Normal University, Wuhan 430079, China; \url{yuyw@ccnu.edu.cn}}
\affiliation{Education Research and Application Center, National Astronomical Data Center, Wuhan 430079, China}
\affiliation{Key Laboratory of Quark and Lepton Physics (Central China Normal University), Ministry of Education, Wuhan 430079, China}

\author[0009-0007-7246-9309]{Ming-Yan Xiao(肖明燕)}
\affiliation{Institute of Astrophysics, Central China Normal University, Wuhan 430079, China; \url{yuyw@ccnu.edu.cn}}
\affiliation{Education Research and Application Center, National Astronomical Data Center, Wuhan 430079, China}
\affiliation{Key Laboratory of Quark and Lepton Physics (Central China Normal University), Ministry of Education, Wuhan 430079, China}

\begin{abstract}

A near-infrared (NIR) excess has been discovered in the emission of the representative fast blue optical transient (FBOT):  AT 2018cow. It was suggested that this NIR excess could be emitted by the dust surrounding the source and, thus, could provide a probe into the nature of its progenitor. We develop a model to describe the influence of the FBOT emission on the environmental dust and, as a result, a dust-free evaporation cavity can be formed on a timescale of one day. Outside this cavity, the surviving dust grains can have different size distributions at different distances to the source. With such a special dust environment, we fit the multi-wavelength light curves of AT 2018cow by taking into account the evolutionary dust echo of the FBOT emission. It is found that the dust temperature can vary with time along with the evolution of the irradiating FBOT emission. Even at a fixed time, the dust temperature can be distributed in a wide range rather than having only a unique value. Furthermore, both the mass of the dust shell and its distance to the FBOT are found to be much larger than those derived with a direct empirical fitting of the NIR spectra but without considering the evolutionary relationship between the spectra.

\end{abstract}


\section{Introduction} \label{sec:intro}
Fast blue optical transients (FBOTs) are a class of unusual transients characterized by their rapid light curve evolution and high brightness in the UV/optical band \citep{drout2014rapidly, inserra2019observational, pasham2022evidence}. The peak absolute magnitudes of FBOTs range from $-16$ to $-23$ and the typical rising time to the maximum brightness is shorter than about ten days \citep{Liu2022}. These characteristics make FBOTs very difficult to be explained by a power of radioactive decays of $^{56}$Ni as normal core-collapse supernovae (CCSNe), because the mass of $^{56}$Ni required by the high luminosity would be usually comparable to and even exceed the total mass of the explosion ejecta inferred from the short timescale. Such a situation also appears in another mysterious phenomenon known as superluminous supernovae (SLSNe), which have been discovered recently too \citep{ofek2007sn,quimby2007sn,smith2007sn,gal2009supernova,Yu2017,gal2019most}. Therefore, analogously to SLSNe, two representative types of models have been put forward in the literature to account for the energy sources of FBOTs. On the one hand, the power of FBOTs could be ascribed to a radiation-mediated shock that propagates into a dense progenitor or circumstellar medium \citep[CSM; e.g.,][]{Chevalier2011,Ginzburg2012,Drout2014,Rest2018,Fox2019,Leung2020,Xiang2021,Pellegrino2022,Gottlieb2022}. In this scenario, to generate a sufficiently dense CSM, the mass loss rates of the progenitors need to be as high as $\sim 1 M_{\odot}$ yr$^{-1}$ \citep{Xiang2021,Pellegrino2022}, which could be however sometimes inconsistent with the radio observations of FBOTs. So, on the other hand, it has been widely suggested that FBOT could be emitted from a low-mass engine-driven ejecta \citep{Yu2015,prentice2018cow,mohan2020nearby,Liu2022,li2024magnetar}. 

Observational support for the engine-powered model may stem from the resemblance of the host galaxies of FBOTs to those of long gamma-ray bursts and hydrogen-poor SLSNe \citep{michalowski2019nature,roychowdhury2019h,lyman2020studying}, as well as from a potential universal correlation among these phenomena, as presented in \cite{liu2022magnetar}. More direct and strong evidence for the existence of the central engine may arise from the multi-wavelength observations of AT 2018cow ranging from radio to gamma-rays \citep{prentice2018cow,rivera2018x,perley2019fast,kuin2019swift,nayana2021ugmrt,ho2022at2018cow}. In particular, a high-amplitude quasi-periodic oscillation with a frequency of 224 Hz was identified in soft X-ray emission \citep{pasham2022evidence}, which can even be regarded as a smoking gun for the connection between the X-ray emission and the remnant compact object. In more detail, the decay of the total luminosity of AT 2018cow, which combines the X-ray and optical emission, can be basically consistent with the spin-down of a magnetar \citep{margutti2019embedded}. Here, the X-ray and optical emission could respectively result from the internal dissipation in the magnetar wind and the subsequent heat of the FBOT ejecta by absorbing the wind emission \citep{Yu2019a,li2024magnetar}. 

In principle, a rapidly rotating magnetar with a low-mass ejecta could originate from various channels, which include core-collapse supernova explosions of ultra-stripped stars \citep{Tauris2013,Tauris2015,Tauris2017,Suwa2015,Moriya2016,Hotokezaka2017,De2018}, electron-capture supernovae \citep{Sawada2022,mor2023bare}, accretion-induced collapse (AIC) of white dwarfs \citep[WD; e.g.,][]{Kasliwal2010,Yu2015,Brooks2017,Yu2019a}, and mergers of double WDs \citep{Yu2019b}, a WD and a neutron star, and even double neutron stars \citep{Yu2013,Zenati2019}. Different channels can lead to different environments for FBOTs.
Therefore, in addition to probing central engines of FBOTs directly, it is expected that their CSM properties can also provide important clues to the nature of their progenitors. Then, the radio emission after FBOTs is usually employed to constrain the density profile of the CSM. Furthermore, an echo of the transient emission could be generated by possibly existing dust in the CSM, which could be discovered from the infrared (IR) observations of FBOTs. Dust grains may condense from an intense CSM that forms from the mass loss of massive stars \citep{dwek1983infrared,andrews2011evidence}. Meanwhile, a super-Chandrasekhar hot WD born from a double WD merger could also generate such a dusty environment \citep{Yu2019b}. In addition, dust grains could also be generated by the explosion ejecta itself, due to its interaction with a dense CSM \citep{pozzo2004source,smith2008sn,chugai2009circumstellar,gall2014rapid} and even just due to the expansion and cooling of the ejecta \citep{kotak2005early,andrews2011photometric,szalai2013twelve}. 
Different dust formation processes may leave different characteristics in the temporal evolution of dust echo emission.

Here we emphasize that a near-IR excess was indeed identified from the observations of AT 2018cow by \cite{perley2019fast}. However, it was initially suggested to arise from a nonthermal emission process, because fitting this near-IR emission directly with a black body led to a temperature higher than the sublimation temperature of dust. Then, \cite{chen2024spectra} ascribed the near-IR emission to the reprocessing of the high-energy emission from the central engine within the ejecta, where the frequency-dependence of the opacity is taken into account. In contrast, \cite{metzger2023dust} argued that dust emission should not be a pure black body. If a modified black body is adopted in the fittings, then the dust temperature can be found to be around a reasonable value of $T\sim 2000$K, which corresponds to the sublimation temperature of carbonaceous grains. However, the spectral fittings presented in \cite{metzger2023dust} were implemented separately for different times without explaining their temporal evolution. Subsequently, an elaborate calculation of the echo light curve has recently been provided by \cite{tuna2025timedependent}, where a torus-shaped dust environment was assumed. In these previous works, the echo was considered to arise only from the first light of FBOTs that destroys the inner CSM dust to form a dust-free cavity. The boundary of the cavity is determined according to the condition that the UV optical depth equals unity. 

However, it should be pointed out that the sublimation of a dust grain is actually dependent on the relationship between the energies it absorbs and emits. Dust destruction can undoubtedly occur in the region where the UV optical depth is less than unity as long as it can receive enough energy. Therefore, the size of the dust-free cavity would be closely related to the duration of the peak emission of FBOTs. In addition, it should also be noted that the surviving dust outside the cavity can be heated persistently by the late transient emission, which thus could provide a remarkable contribution to the echo emission at late times. Therefore, in this work, we revisit the calculation of the dust echo of FBOTs by taking into account the above-mentioned effects and fitting the spectra and multi-wavelength light curves of AT 2018cow simultaneously.

%%%%=================================================%%%%%
\section{The Evaporation and emission of dust}\label{sec:Evaporation}
In our calculations, a dust shell is assumed to pre-exist surrounding the progenitor of FBOT, which have a density profile as
\begin{equation}
n (r) = n_{\rm d} \left(\frac{r}{R_{*}}\right)^{-3}, {~~\rm for~}R_{\rm in}\leq r\leq R_{\rm out},\label{Eq:density}
\end{equation} 
where the power-law index $-3$ is adopted according a previous work of \cite{bright2022radio} and $n_{\rm d}$ is the number density of dust grains at the reference radius of $R_{*}\sim 10^{16}$~cm. At each radius of the dust shell, it is further assumed that the dust grains distribute on their size $a$ as a function of \citep{gall2014rapid}
\begin{equation}
 f (a) \propto a^{- \alpha}, 
\end{equation}
which is normalized to be unity in the interval $[a_{\min}, a_{\max}]$ as $\int_{a_{\min}}^{a_{\max}} f (a) \text{d} a = 1$ and the power-law index $\alpha$ typically ranges from 2.0 to 4.5. 

When an FBOT occurs, the luminous emission would irradiate the surrounding dust grains and provide them energy at a rate of \citep{draine1984optical,laor1993spectroscopic}
\begin{equation}
 H = \frac{\pi a^2 R_{\rm{ph}}^2}{r^2} \int \pi B_{\nu} (T_{\rm{ph}}) Q_{\rm{abs}, \nu} (a) \text{d} \nu,\label{Eq:heat}
\end{equation}
where $R_{\tmop{ph}}$ and $T_{\tmop{ph}}$ are the radius and temperature of the photosphere of the transient emission, respectively, and $B_{\nu}$ is the intensity of a black body.  The value of the absorption coefficient $Q_{\rm{abs}, \nu} (a)$ can be found from Draine's website\footnote{\url{https://www.astro.princeton.edu/~draine/dust/dust.diel.html}} for different sizes of grains over the wavelength range of $10^{-3}-10^{3} \, \mu m$. As a result of the heating effect, the temperature of dust grains, $T_{\rm d}$, can be determined by the following energy conservation law: 
\begin{equation}\label{Eq:dU_dt}
 m C_{\rm m} \frac{\text{d} T_{\rm d} }{\text{d} t}=H+{E_{\rm b}\over m_{a}} \frac{\text{d} m}{\text{d} t} - \ell _{\tmop{emit}},
\end{equation}
where $m$ is the mass of a dust grain corresponding to the radius $a$, $C_{\rm m}$ is the specific heat capacity for temperature above the Debye temperature, $E_{\rm b}$ is the binding energy per atom in the dust lattice, and $m_{a}$ is the mass of the atom. The power of the thermal emission of the dust grain can be given by
\begin{equation}\label{Eq:Lemit}
\ell _{\rm{emit}} = 4 \pi a^2 \int \pi B_{\nu} (T_{\rm d}) Q_{\tmop{abs}, \nu} (a)  \text{d} \nu.
\end{equation} 
As a result of the irradiation by the FBOT emission, the dust grains can not only be heated but can also evaporate at a rate of
\begin{equation}\label{Eq:dmdt1}
\frac{\text{d} m }{\text{d} t} = 4\pi a^2\rho{\text{d}a\over \text{d}t}, 
\end{equation}
with \citep{waxman2000dust}
\begin{equation}\label{Eq:dadt}
\frac{\text{d} a }{\text{d} t} = - \zeta \exp\left({ - {E_{\rm b}\over k_{\rm B} T_{\rm{d}}}}\right).
\end{equation}
In the following calculations, we would only take into account graphite dusts but ignore possible silicates. Then, the proportional coefficient in the above equation can be taken as $\zeta=4.13\times10^{6}\ \text {cm s}^{-1}$ \citep{waxman2000dust} and the other related parameters are given as follows: $C_{\rm m}= 2.1\times10^{7} \rm ~erg~ g^{-1} K^{-1}$, $E_{\rm b}=7.5$ eV, $m_{a}=1.99\times10^{-23}$ g, $\rho=2.26 ~\rm g ~cm^{-3}$, and $\alpha=3.5$ \citep{mathis1977size,draine1984optical,jiang2021infrared}. Here, the values of parameters such as $\alpha$ are actually referenced from the ISM situation, which may not truly be applicable to FBOTs. However, in order to minimize the degrees of freedom of the model as much as possible, and considering that the ISM environment is also the result of contributions from various explosive phenomena, we still adopted these values instead of treating them as free parameters.

\begin{figure}[htbp!]
    \centering
    \includegraphics[scale = 0.25]{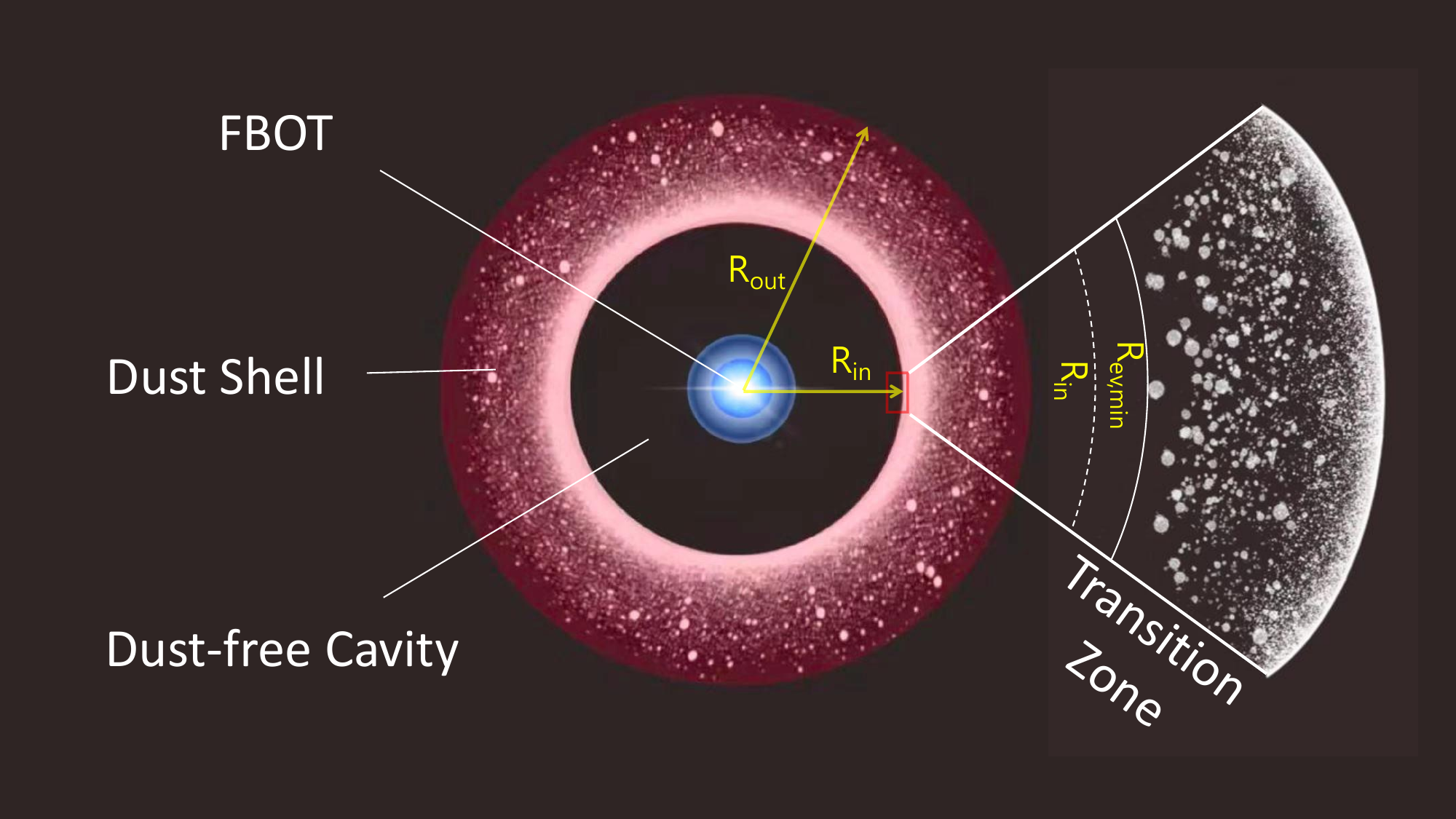}
    \caption{The schematic diagram of the model (not to scale).}
    \label{Fig.dust_fig}
\end{figure}

Determined by Equations (\ref{Eq:heat}-\ref{Eq:dadt}), the size of dust grains can be reduced quickly, especially, when its temperature is very close to the sublimation temperature. Therefore, at sufficiently small distances to the FBOT, the heating of the dust grains can be very effective and thus all dusts there would be evaporated completely. As illustrated in Figure \ref{Fig.dust_fig}, a dust-free evaporation cavity can form surrounding the FBOT source, in which no dust can survive. In contrast, outside the cavity, the farther away from the center, the more dust can survive. 
Here, let us estimate the absorption depth of a dust shell prior to dust destruction, which can be expressed as
\begin{eqnarray}
\tau_{\rm abs} &=&\int\int \bar{Q}_{\rm abs}\pi a^2 f(a)n(r)  da dr\nonumber\\
&\approx&{\pi (\alpha-1)\bar{Q}_{\rm abs}a_{\min}^2n_{\rm d}R_*^3\over 2(\alpha-3)R_{\rm in}^2},\nonumber\\
&\approx&0.8\left({n_{\rm d}\over 0.01{\rm cm^{-3}}}\right)\left({R_{\rm in}\over 10^{16} {\rm cm}}\right)^{-2},
\end{eqnarray}
where $\bar{Q}_{\rm abs}\pi a^2$ represents the effective cross section of a dust grain and $\bar{Q}_{\rm abs}\sim 0.1$ \citep{draine1984optical} is adopted. This indicates that an FBOT emission could be blocked by a dust shell if its inner radius is much smaller than $10^{16}$ cm and the density profile described by Eq. (\ref{Eq:density}) can indeed extend to such small radii. By contrast, for $R_{\rm in}\gtrsim 10^{16}$ cm, the FBOT emission can successfully emerge from the dust shell to be detected and the AT 2018cow event should be the case. Furthermore, a transition zone can be formed outside the dust-free cavity, where the minimum dust grains are destroyed but the maximum ones survive. This transition zone would play the most important role in determining the echo emission of the FBOT, since the dust in it are the closest ones to the FBOT emission. Therefore, it is of fundamental importance to describe the size distribution of the dust grains in this transition zone.

%%%%=================================================%%%%%

\section{Application to AT 2018cow} \label{sec:fit}
\subsection{FBOT Emission}\label{sec:Mag}
Since the focus of this paper is on the dust echo of FBOTs, the calculation of the FBOT emission would be simply implemented by according to the analytical solution of \cite{Arnett1982}. Specifically, the bolometric luminosity of the FBOT emission is given by: 
\begin{equation}
\begin{split}
L_{\text{rad}}(t) &= e^{-\left(\frac{t}{t_{\text{diff}}}\right)^2} \int_0^t 2 L_{\text{sd}}(t')\left(1 - e^{-\tau_{\rm{X}}}\right) \\
&\quad \times \frac{t'}{t_{\text{diff}}} e^{\left(\frac{t'}{t_{\text{diff}}}\right)^2} \frac{\text{d}t'}{t_{\text{diff}}}
\end{split}
\label{Eq:Lrad}
\end{equation}
with
\begin{equation}
t_{\text{diff}} \equiv \left( \frac{3\kappa M_\mathrm{ej}}{4 \pi v_\mathrm{ej} c} \right) ^{1/2}
\end{equation}
being the photon diffusion timescale of the FBOT ejecta and 
\begin{equation}
\tau_{\rm{X}} \equiv  \frac{3\kappa_{\rm X} M_\mathrm{ej}}{4 \pi v_\mathrm{ej} ^{2}t ^{2}} 
\end{equation}
is the optical depth for X-ray photons, representing the absorption of the magnetar wind emission by the FBOT ejecta. Here, $\kappa$ and $\kappa_{\rm X}$ are respectively the opacities of the ejecta material in the optical and X-ray bands and the magnetar wind emission is assumed mainly in X-rays. $M_{\rm ej}$ and $v_{\rm ej}$ are the mass and expansion velocity of the ejecta. In Equation (\ref{Eq:Lrad}), we assume the energy supply of the FBOT emission is contributed by the spin-down of a rapidly rotating magnetar and express its luminosity as usual as \citep{dai1998gamma,kasen2010supernova,shapiro2024black}

\begin{equation}
L_{\rm sd}(t) \;=\; L_{\rm sd,i}\,\biggl(1 + \frac{t}{t_{\rm sd}}\biggr)^{-2},
\end{equation}
where the initial value of the spin-down luminosity and the spin-down timescale are given by $L_{\rm sd,i} =  10^{47}\,{\rm erg}\,{\rm s}^{-1}\,P_{\rm i,-3}^{-4}B_{\rm p,14}^2$ and $t_{\rm sd} \simeq 2\times10^5\,{\rm s}\,P_{\rm i,-3}^2B_{\rm p,14}^{-2}$, respectively, with $P_{\rm i}$ and $B_{\rm p}$ being the initial spin period and the dipolar magnetic strength of the magnetar. 
%depends on the spin period ($P_{\rm i}$) and magnetic field strength ($B_{\rm p}$) of the magnetar as usual  

In order to calculate the multiband light curves of the FBOT, we assume a black body spectra as usual and determine its temperature by
\begin{equation}
T_\mathrm{ph}(t) \equiv \mathrm{max}\left\{\left[\frac{L_{\mathrm{rad}}(t)}{4 \pi \sigma R_\mathrm{ph}^2}\right]^{1/4},T_\mathrm{floor}\right\}\label{Eq:Tem}
\end{equation}
where $\sigma$ is the Stefan$-$Boltzmann constant and $R_{\rm ph}$ is the photospheric radius of the emitting ejecta. Before the ejecta becomes transparent, the photospheric velocity can be considered to approximately equate the expanding velocity of the ejecta and thus we have $R_{\rm ph}\approx v_{\rm ej}t$. By contrast, at late times, we would take the temperature as a constant ($T_\mathrm{floor}$ ), which is widely assumed in modeling supernova-like transients \citep[e.g.,][]{nicholl2017magnetar,liu2022magnetar} and taken as a free parameter here. In this case, the photospheric radius is given by 
\begin{equation}
R_{\rm ph}\approx \left(\frac{L_{\mathrm{rad}}}{4 \pi \sigma T_{\rm floor}^4}\right)^{1/2}. 
\end{equation}

It should be noticed that the above physical assumptions are actually not prerequisite for the following calculations of dust echo. The purpose of introducing above model is to provide incident lights for the dust echo that can be consistent with the multi-wavelength observations. In principle, we could also directly use empirical functions to describe these incident lights.

%%%%=================================================%%%%%

\begin{figure*}[htbp!]
    \centering
    \includegraphics[width=0.45\textwidth]{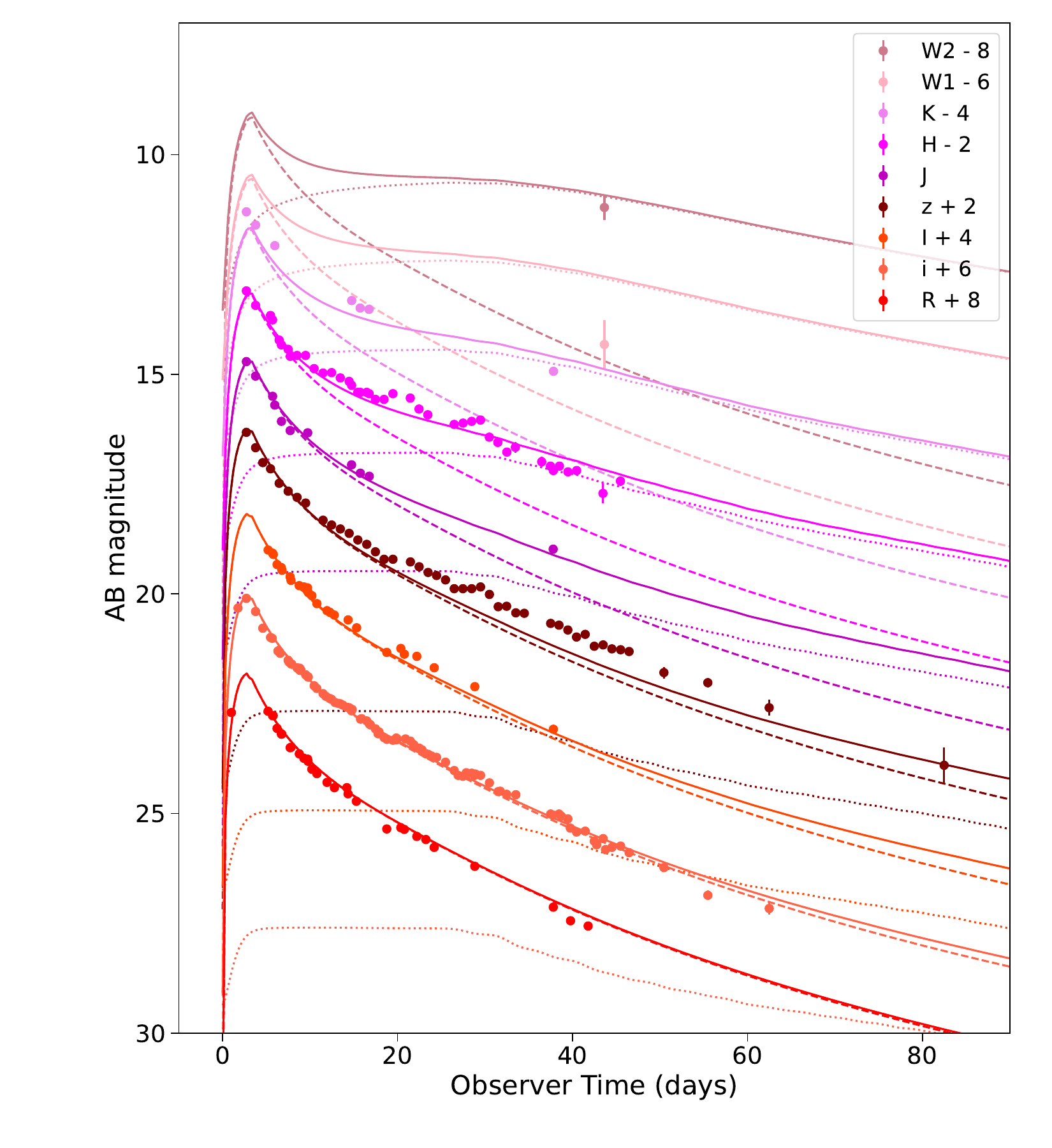}
    \includegraphics[width=0.45\textwidth]{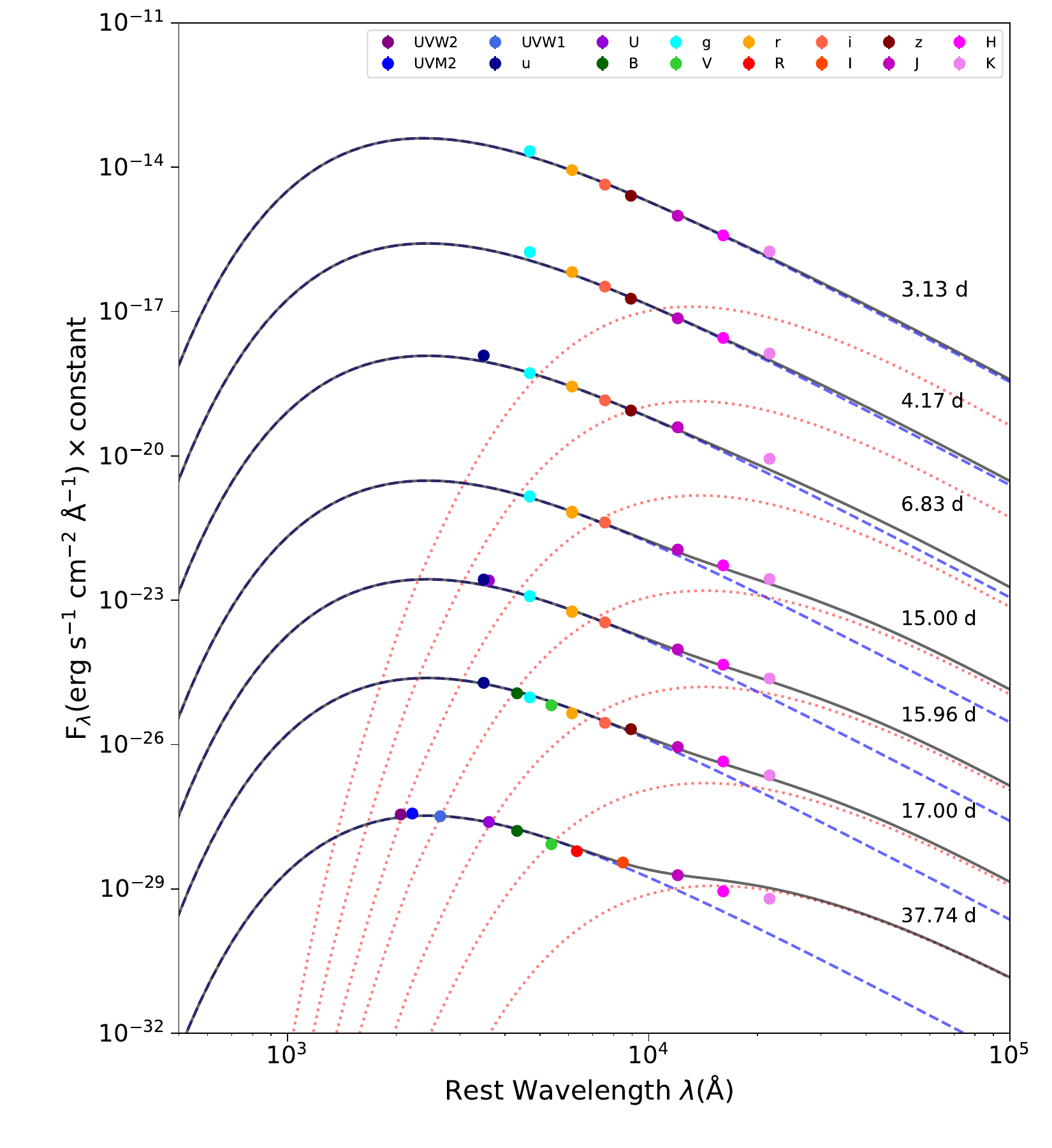}
    \caption{{\it Left}: Fitting to the multi-wavelength light curves of AT 2018cow. {\it Right}: A comparison between the model-predicted spectra and observational ones at seven time points. The dashed and dotted lines represent the FBOT and dust echo emission, respectively. }
    \label{Fig.LC_SED}
\end{figure*}

\begin{table}[htbp!]
\centering
\caption{Free Parameters, Priors, and Best-fitting Results}
\begin{tabular}{cccc} 
\hline
Parameter & Prior & Allowed Range & Posterior \\
\hline
$M_{\rm ej}/M_{\odot}$ & Flat & $[0.01,2]$ & $0.03_{-0.01}^{+0.01}$ \\
$P_{\rm i}/{\rm ms}$ & Flat  &$[0.1,20]$ & $16.71_{-0.31}^{+0.33}$ \\
$\mathrm{log}_{10}(B_{\rm p}/10^{14}{\rm G})$ & Log-flat  & $[-2,2]$ & $1.32_{-0.00}^{+0.00}$ \\
$v_{\rm ej}/10^{9}{\rm cm}\,{\rm s^{-1}}$ & Flat  & $[0.1,10]$ & $7.41_{-0.24}^{+0.27}$ \\
$\mathrm{log}_{10}(\kappa_{\rm X}/{\rm cm}^{2}\,{\rm g}^{-1})$ & Log-flat  & $[-1,2]$ & $1.59_{-0.09}^{+0.13}$ \\
$\kappa/{\rm cm}^{2}\,{\rm g}^{-1}$ & Flat  & $[0.17,0.34]$ & $0.23_{-0.04}^{+0.07}$\\
$T_{\rm floor}/10^{3}\,{\rm K}$ & Flat  & $[3,100]$ & $11.62_{-0.03}^{+0.03}$\\
$t_{\rm shift}/{\rm days}$ & Flat  & $[0.01,1.31]$ & $1.01_{-0.05}^{+0.05}$ \\
\hline
$\mathrm{log}_{10}(n_{\rm d}/{\rm cm}^{-3})$ & Log-flat  & $[-3.5,0]$ & $-2.37_{-0.01}^{+0.00}$ \\
$\mathrm{log}_{10}(R_{\rm out}/{\rm cm})$ & Log-flat  & $[16.9,17.9]$ & $17.77_{-0.01}^{+0.01}$ \\
\hline
\end{tabular}\label{Tab:para}
\end{table}

\subsection{Dust Echo}\label{sec:echo_modle}

The echo of FBOT emission should be delayed by a time relative to the FBOT itself, because the radiation path through the dust is longer than the direct one from the central source \citep{dwek1983infrared,pearce1984infra,dwek1985infrared}. The delayed time can be expressed as
\begin{equation}
\tau_{\rm d}  =\frac{r}{c}(1-\cos \theta),
\end{equation}
which depends on the angle $\theta$ of the direction of the dust grain relative to the line of sight. Then, for a dust shell spreading from $R_{\rm in}$ to $R_{\rm out}$, its echo emission luminosity detected by observers at a time $t$ can be calculated by the following integration: 
\begin{eqnarray}\label{Eq:Fv_echo}
L_\nu^{\text{Echo}}(t) &=& 2 \pi c \int_{R_{\rm in}}^{R_{\rm out}}\int_0^{\min[2r/c,t]} \int_{a_{\min}}^{a_{\max}} 
 \nonumber\\
&&\times\ell _{\rm emit,\nu}(t') f(a) \, n(r) r\text{d}a \,      \text{d} \tau_{\rm d} \text{d} r,
\end{eqnarray}
where the differential volume of the dust shell is expressed by $\text{d} V =2 \pi r \sin \theta {\rm d} \theta \text{d} r= 2 \pi r c  \text{d} \tau_{\rm d} \text{d} r $ and
\begin{equation}\label{L_nu_a}
\ell _{\rm emit,\nu} = 4 \pi a^2 \pi B_{\nu} (T_{\rm d}) Q_{\rm{abs}, \nu} (a) 
\end{equation}
is the radiative luminosity of a dust grain at the time $t'=t-\tau_{\rm d}$. Here, it should be noted that the actual values of $R_{\rm in}$ and $a_{\rm min}$ would finally be determined by dust evaporation due to FBOT emission, which could be very different from their initial values. In addition, besides the thermal emission, the dust grains can also scatter the incident FBOT lights into all directions including the direction of the observer \citep{laor1993spectroscopic}. Then, this scattering effect can in principle contribute to the delayed echo emission too. However, the scattering would not change the energy of the incident photons and thus cannot influence the concerned IR emission. Furthermore, since the delay time is not longer than the duration of the FBOT emission, the FBOT emission can always outshine the scattering component and make it negligible.

\subsection{Fitting results and analyses}\label{sec:lcfit}

With the model presented above, we can calculate the fluxes of FBOT emission and its echo by the following two formulae:
\begin{equation}
 F_{\nu}^{\tmop{FBOT}} = \frac{\pi B_{\nu}
   (T_{\tmop{ph}}) R_{\tmop{ph}}^2}{D^2}
\end{equation}
and
\begin{equation}
 F_{\nu}^{\tmop{Echo}} = \frac{L_{\nu}^{\rm Echo}}{4 \pi D^2},
\end{equation}
where $D$ is the luminosity distance of the FBOT. Here, it should be noticed that the zero time of the FBOT explosion can be earlier than the achievement of the first observation and, thus, a shift $t_{\rm shift}$ is added to the time of all data in our fitting of the multi-wavelength light curves of AT 2018cow. The data presented in Figure \ref{Fig.LC_SED} are taken primarily from \cite{perley2019fast}, while the two points at W1 ($3.4\mu$m) and W2 ($4.6\mu$m) bands are obtained with the Wide-field Infrared Survey Explorer \citep[WISE,][]{wright2010wide} and its extended mission, the Near-Earth Object WISE Reactivation ~\citep[NEOWISE,][]{mainzer2014initial}\footnote{NEOWISE had visited the position of AT2018cow at MJD=58328, which is 41 days after the optical peak. We have taken the W1 and W2 data from Ning Jiang through private communication. He performed point spread function (PSF) photometry on the difference images of time-resolved NEOWISE coadds~\citep{meisner2018time} with the first NEOWISE epoch (MJD~56886) as a reference following~\citet{jiang2021infrared} and found a clear W2 excess with a flux of $0.076\pm0.017$~mJy and a tentative W1 excess with a flux of $0.027\pm0.011$.}.

\begin{figure*}[htbp!]
    \centering
    \includegraphics[width=0.99\textwidth]{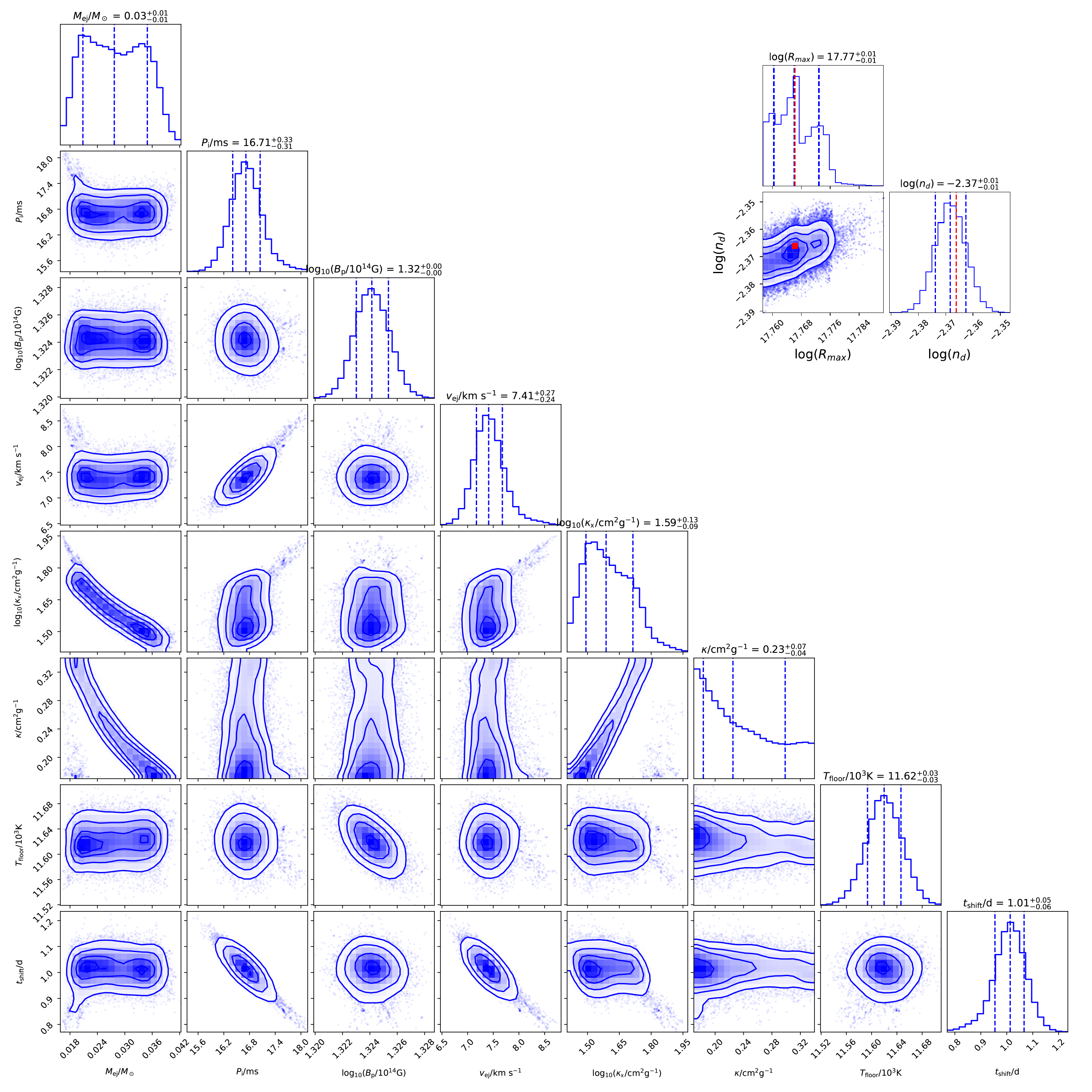}
    \caption{Parameter posteriors of the fittings of AT 2018cow.}
    \label{Fig.corner}
\end{figure*}

Our fitting of the multi-wavelength light curves of AT 2018cow is conducted in two steps. First, we confront the FBOT model with the data of the light curve in the $I, i, R, r, V, g, B, U, u$ bands without considering the dust echo. Second, with best-fit parameters derived from the first step, we model the light curves in $z, J, H, K, W1$, and $W2$ bands by including the dust echo emission. The best-fit result is presented in the left panel of Figure \ref{Fig.LC_SED}. The constraints on the model parameters are
obtained with the {\it emcee} package \citep{foreman2013emcee}. The corner plots of the posterior parameter regions are presented in Figures \ref{Fig.corner} and the $1-\sigma$ posterior ranges are listed in Table \ref{Tab:para} as well as their priors. The fitting result demonstrates that the invoking of the dust echo can provide a general explanation of the emission excesses in the $K, H, J$ bands\footnote{The remaining excess in the $z$ band from 20 to 60 days may indicate that a more realistic spectrum could somewhat deviate from a blackbody}. A comparison between the model-predicted spectra and the observational ones is also presented in the right panel of Figure \ref{Fig.LC_SED}, further demonstrating the applicability of the dust model in explaining the near-IR excess of AT 2018cow. 

\begin{figure*}[htbp!]
    \centering
    \includegraphics[width=0.5\textwidth]{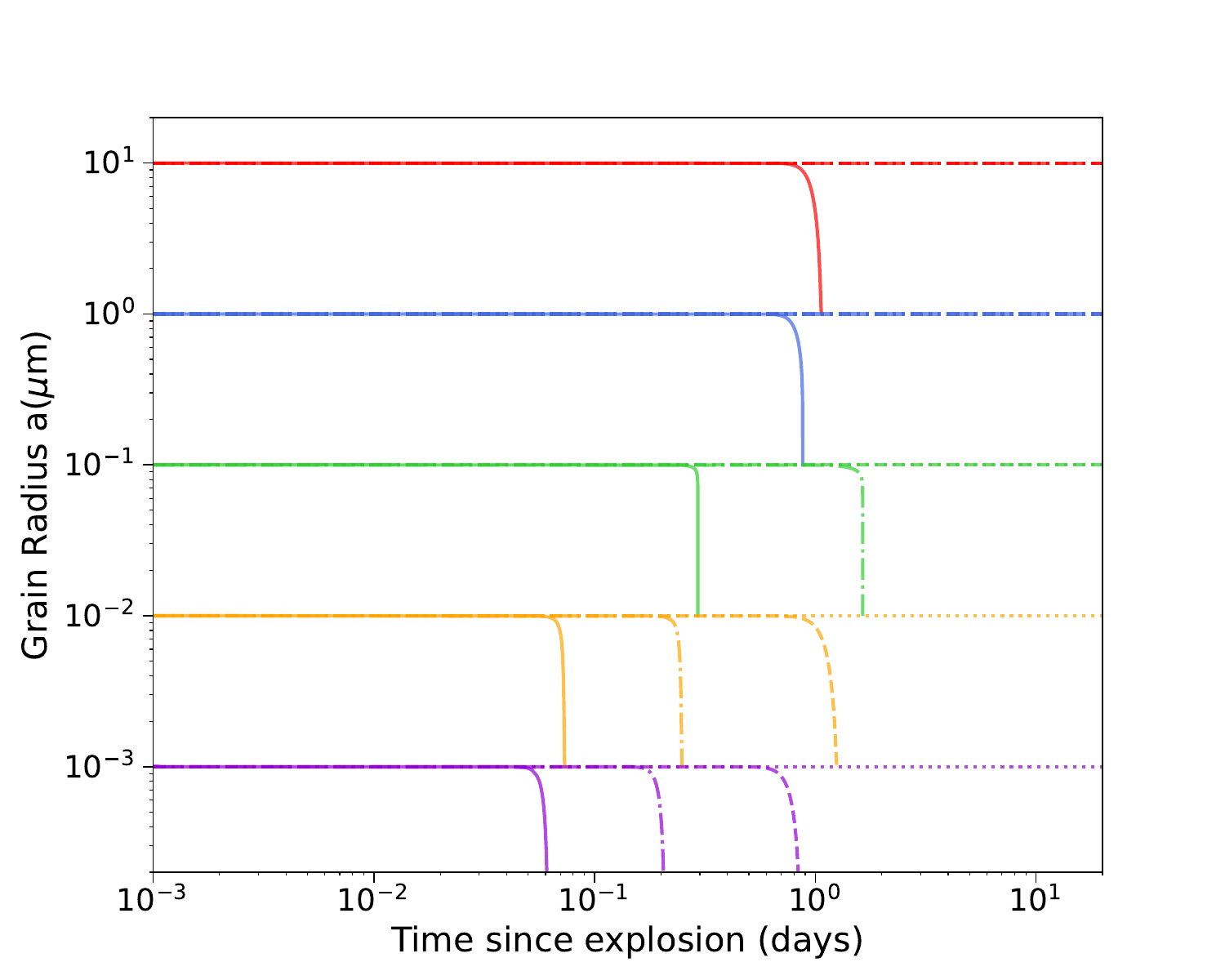}\includegraphics[width=0.5\textwidth]{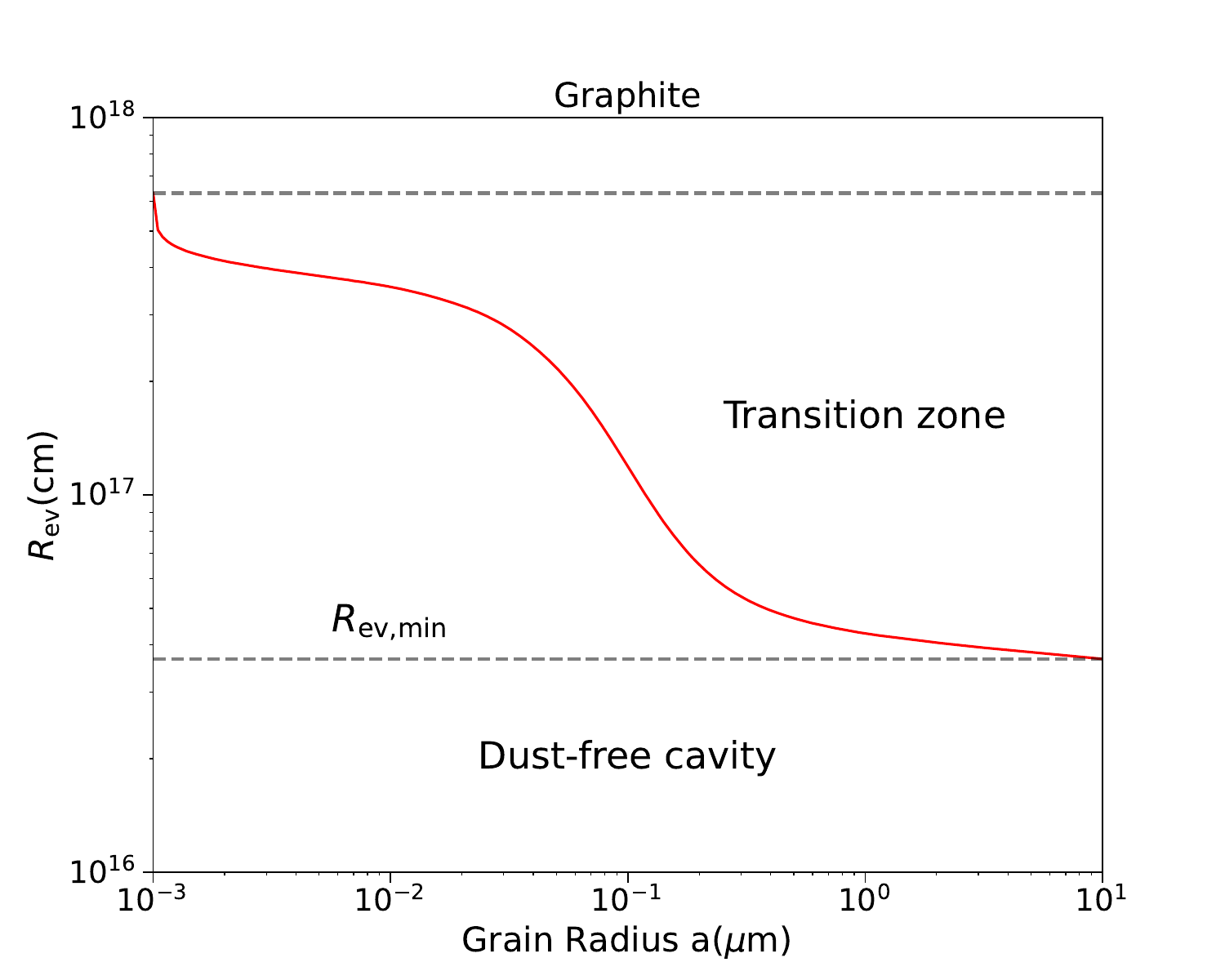} %\includegraphics[width=0.50\textwidth]{Td_t.pdf}
    \caption{{\it Left}: The size evolution of dust grains for different initial sizes. The solid, dotted-dashed, dashed, and dotted lines represent the distances to the FBOT source of $2\times10^{16}$ cm, $8\times10^{16}$ cm, $3\times10^{17}$ cm, and $1\times10^{18}$ cm, respectively. {\it Right}: The evaporation radius for different size of dust grains, inside which the dusts of a size less than $a$ can be evaporated completely within a short time.}
    \label{Fig.a_t_Revap}
\end{figure*}

In our calculations, the initial values of $a_{\min}$ and $a_{\max}$ are taken to be 0.001$\mu m$ and 10$\mu m$, respectively. The size evolutions of different dust grains at different distances are presented in the left panel of Figure \ref{Fig.a_t_Revap}. It is shown that, at the distance of $2\times10^{16}$ cm, all dust grains can be destroyed completely within 1 day after the FBOT occurs. The lifetime of the dust grains can be prolonged as the distance to the FBOT increases. The larger the grain size, the longer the grain's lifetime. For a specific dust size, its corresponding evaporation radius $R_{\rm ev}$ can be roughly determined by the condition that the dust temperature reaches its sublimation value at which the energy absorption and emission of the dust \citep{metzger2023dust}
\begin{equation}
{L_{\rm p}\pi a^2Q_{\rm abs,UV}\over 4\pi R_{\rm ev}^2}\sim 4\pi a^2\sigma T_{\rm ev}^4Q_{\rm abs,IR},
\end{equation}
yielding
\begin{eqnarray}
R_{\rm ev}&\sim&{1\over4}\left[{L_{\rm p}Q_{\rm abs,UV}\over \pi \sigma T_{\rm ev}^4Q_{\rm abs,IR}}\right]^{1/2}. 
\end{eqnarray} 
The specific dependence of $R_{\rm ev}$ on $a$ is determined by the $a-$dependence of $Q_{\rm abs,UV}$ and $Q_{\rm abs,IR}$ and the numerical result is shown in the right panel of Figure \ref{Fig.a_t_Revap} by the solid line. For distances smaller than $\sim 10^{18}$ cm, the actual minimum size of the dust grains can become obviously larger than its initial value and even larger than the value of $a_{\max}$ for $r<2\times10^{16}$. So, the initial value of the inner radius of the dust shell, $R_{\rm in}$, is basically irrelevant to our calculations of the echo emission. The property of the echo emission is mainly determined by the dusts outside the evaporation cavity. In the boundary region of the cavity, the closer the center is, the more important is the influence of the dust. 
%The calculation result is not very sensitive to the value of $R_{\rm out}$.  

\begin{figure}[htbp!]
    \centering
    \includegraphics[width=0.5\textwidth]{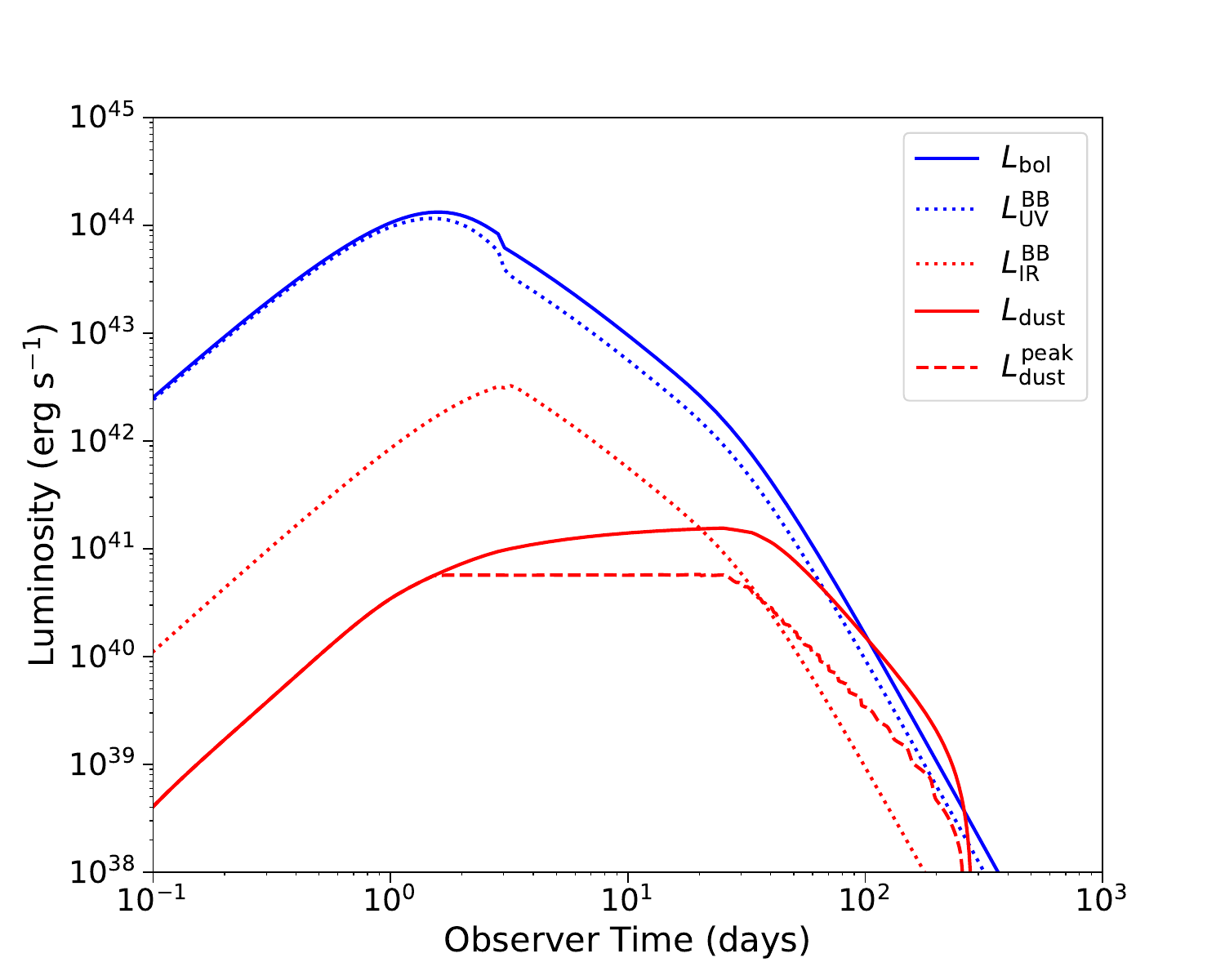}
    \caption{The bolometric light curves of the FBOT (solid blue) and dust echo (solid red) emission for parameters derived from the fitting of AT 2018cow. The dotted lines show the UV ($10^{15-17}$Hz) and IR ($10^{11-13}$Hz) components of the FBOT. The dashed line is the echo of the FBOT emission only before its peak.}
    \label{Fig.LC}
\end{figure}

For a more direct impression of the relationship between the FBOT and echo emission, we display their bolometric light curves in Figure \ref{Fig.LC}, where the FBOT light curve can be basically described by a $t^{2}$ rise and a $t^{-2}$ decline around the peak because the spin-down timescale is just comparable to the diffusion timescale \citep{Yu2015}. The slight break in the light curves appearing around 2-3 days is due to the transition of the two different definitions of the emission temperature as presented in Eq. (\ref{Eq:Tem}). Initially, the echo emission can rise accompanying with the increase of the irradiated FBOT emission. Here, if we define the FBOT emission during its rising stage as the first light and only consider its echo, then the echo emission (dashed line) would quickly enter into a plateau state for a period of $2R_{\rm ev,min}/c$. After that, the echo emission decreases as $\sim t^{-3}$ as the final light transferring from the inner to the outer part of the dust shell. Nevertheless, in fact, the dust can actually not only be heated by the first light of the FBOT, but could be continuously heated even though the FBOT emission enters into the decay phase. So, the usually predicted echo plateau could actually be replaced by a gentle growth, leading to an IR flux several times higher than that obtained by only considering the echo of the first light.

%\begin{figure}[htbp!]
%    \centering
%    \includegraphics[width=0.5\textwidth]{tau_UV.pdf}
%    \caption{Evolution of the absorption optical depth in the UVW2 band of the dust environment during the evaporation process.}
%    \label{Fig.dust_tau}
%\end{figure}

\subsection{Comparison with previous results}

\begin{figure}[htbp!]
    \centering
    \includegraphics[width=0.50\textwidth]{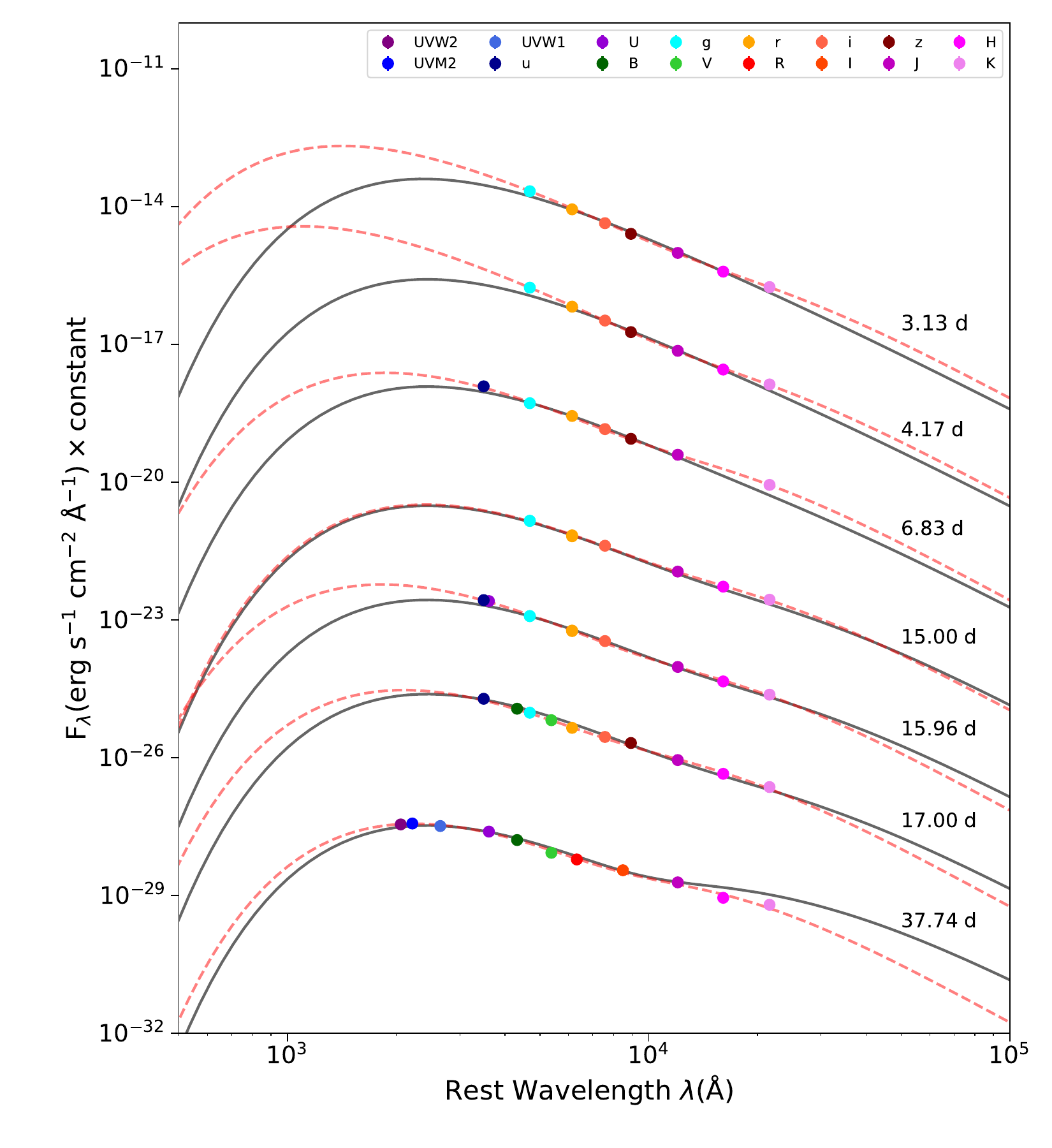}
    \caption{Independent fittings for the spectra of AT 2018cow at seven epochs (dashed lines), in comparison with our fittings (solid lines) with the evolutionary dust echo model. }
    \label{Fig.Spect2}
\end{figure}

\begin{figure*}[htbp!]
    \centering
    \includegraphics[width=0.32\textwidth]{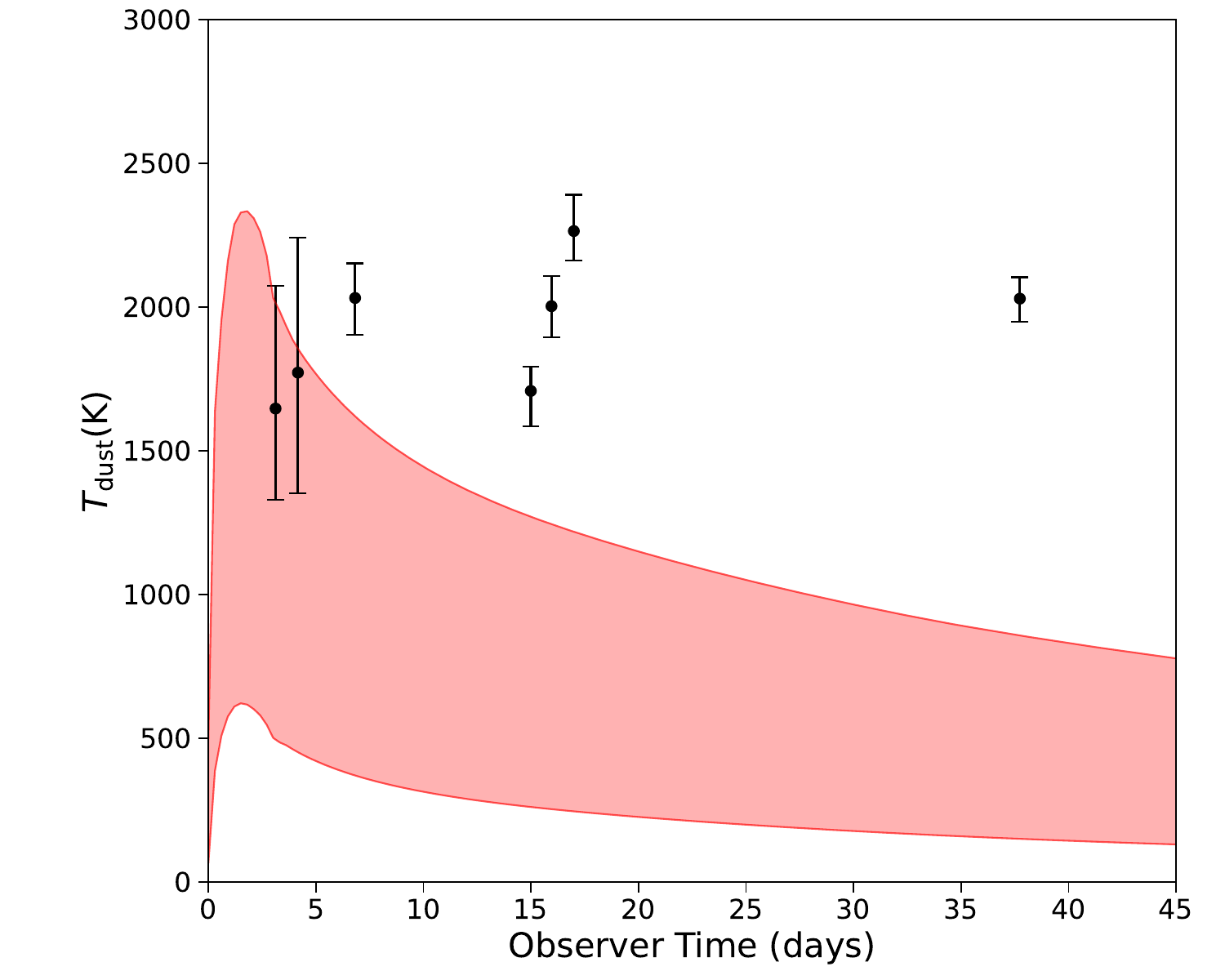} \includegraphics[width=0.32\textwidth]{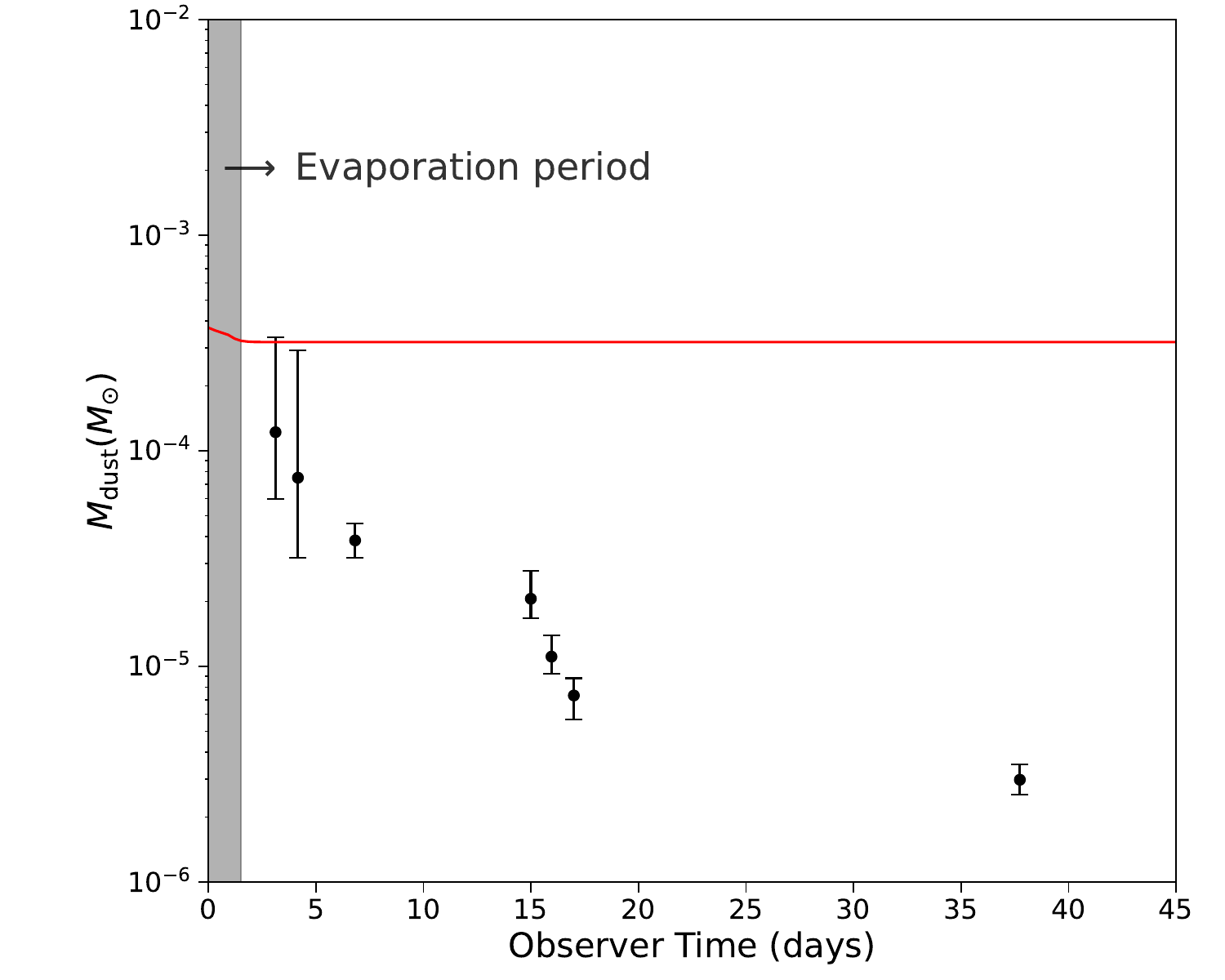}
    \includegraphics[width=0.32\textwidth]{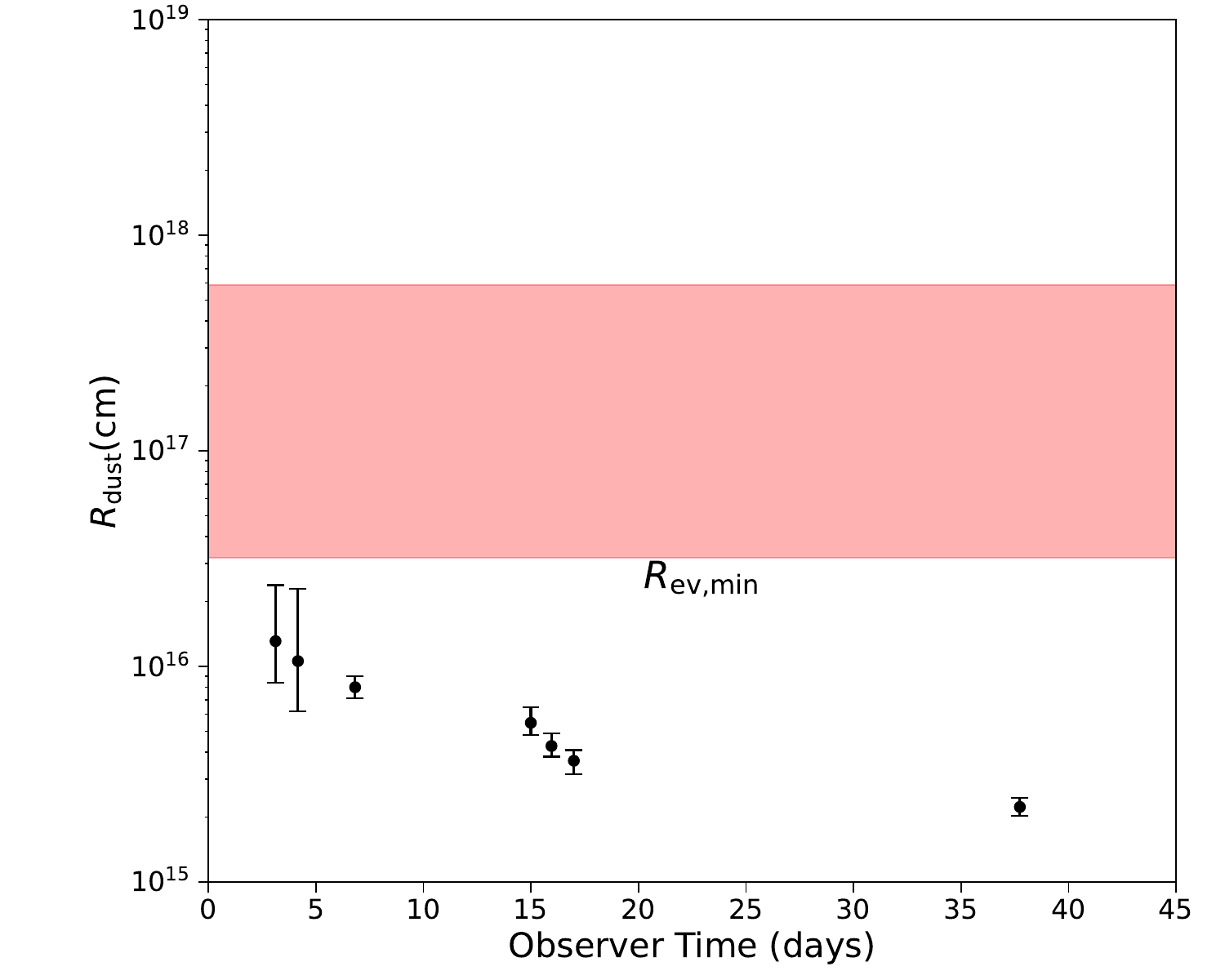}
    \caption{The dust parameters (solid circles) for the spectral fittings of AT 2018cow for seven different times as shown in Figure \ref{Fig.Spect2}, where the fittings are implemented without considering the physical connection between the spectra. For comparison, the corresponding parameter values obtained in our evolutionary dust echo model are shown by the red lines or bands, where the bands indicate that the parameters actually distribute in a range rather than being a fixed value. The dust evaporation occurs in the first day where the dust mass decreases a bit.}
    \label{Fig.DustPara}
\end{figure*}

For comparison, we also fit the AT 2018cow spectra separately and independently, as usually did in the previous literature \citep[e.g., ][]{fox2010disentangling,fox2011spitzer,sarangi2018delayed,li2022using}, where the self-consistency and evolutionary relationship of the model parameters, as well as the time delay of the echo, are not taken into account. The fitting results are presented in Figure \ref{Fig.Spect2} and the values of the parameters $M_{\rm dust}$, $T_{\rm dust}$, and $R_{\rm dust}$ are shown in Figure \ref{Fig.DustPara}. Here, $M_{\rm dust}$ is the total mass of the dusts and the distribution of this mass as well as the thickness of the dust shell are ignored. Then, the effective radius of the dust shell is defined by
\begin{equation}\label{Eq:Rd_Md}
R_{\rm d,eff} = \left( \frac{L_{\rm d}}{4 \pi \sigma T_{\rm d}^4} \right)^{1/2}.
\end{equation}
Figure \ref{Fig.DustPara} shows that the parameter values for the different spectra are very different. However, both the mass and the radius of the dust shell can vary significantly on a timescale of several tens of days, which is difficult to understand. There is no physical explanation to explain why the dust mass can decrease from $\sim10^{-4}M_{\odot}$ to $\sim10^{-6}M_{\odot}$ in several decades of days as the radius of the dust shell decreases with time, as shown by the red band in Figure \ref{Fig.DustPara}. The appearance of these nonphysical results is because the dust temperature is always derived to be around 2000 K. However, this is very likely not to be true, because the echo spectra could actually be a multi-temperature blackbody. As discovered in our evolutionary echo model, even at a fixed time, the dust temperature should be distributed in a wide range because, under the same FBOT irradiation, the dust grains of smaller sizes can be heated more effectively. Furthermore, the dust temperature would decrease with time as the FBOT irradiation weakens. After taking these temperature distribution and evolution into account, the dust mass and radius can be found to remain constant except for the first days during which the initial evaporation occurs, as also shown in Figure \ref{Fig.DustPara}. In our model, the total mass of the dust shell is calculated by
\begin{eqnarray}
     M_{\rm d} (t) &=& \frac{16\pi^2}{3}\rho\int_{R_{\rm ev,min}}^{R_{\rm out}} \int_{a_{\min}(r)}^{a_{\max}}  a^3 f (a) n (r)r^2 \text{d} a \text{d} r\nonumber\\
     &\approx&\frac{16\pi^2 (\alpha-1)n_{\rm d}R_{*}^3\rho a_{\max}^3}{3(4-\alpha)}\left({a_{\max}\over a_{\min}}\right)^{1-\alpha}\ln\left({R_{\rm out}\over R_{\rm ev,min}}\right)\nonumber\\
      &\approx& 3.0\times10^{-4}M_{\odot}\left({n_{\rm d}\over 0.01{\rm cm^{-3}}}\right)\ln\left({R_{\rm out}\over R_{\rm ev,min}}\right).
\end{eqnarray}
In the above equation, the lower limit of the inner integral, $a_{\min}(r)$, should be a function of $r$ that is determined by the evaporation process, as shown by the transition zone in Figure \ref{Fig.a_t_Revap}. However, since $a_{\max}\gg a_{\min}$, the integral result is actually not sensitive to the $r$-dependence of $a_{\min}$ and, thus, we can get the approximated analytical expression.  

%{\color{red} Based on statistical analyses of dust masses in numerous SNe\footnote{\url{https://nebulousresearch.org/dustmasses/}}, fitting SEDs constrains the dust mass within the first 100 days after explosion to be on the order of $10^{-5}-10^{-3} M_{\odot}$.}

\section{Summary and discussion} \label{sec:conclusions}
The progenitors of mysterious FBOTs could experience a significant mass loss in their late evolution, leading to the formation of particular CSM environment for the final explosion event. Through cooling and condensation of the CSM, a remarkable amount of dust could be formed in a range of distances to the progenitor and the size of the dust grains could also be distributed in a range typically from 0.001$\mu$m to 10$\mu$m. The existence of such a dust shell can lead to the appearance of an echo of the FBOT emission, which can be primarily in the IR band. So, an IR excess in FBOT emission is expected to be a probe to its dusty environment and such a candidate was indeed discovered and analyzed in the AT 2018cow event. However, in previous works, the properties of the dust environment including the temperature, mass, and distance to the central source are usually derived by empirically fitting the IR spectra but without considering the physical evolution of the system. 

In this work, we develop a model to evaluate the influence of the FBOT emission on the dust environment, which can lead to the formation of a dust-free evaporation cavity. %, i.e., the inner radius of the dust shell is actually determined by the evaporation effect. 
Outside the cavity, the size distribution of dust grains is further dependent on their distance to the cavity because in such a transition zone, the dust is partially destroyed. Taking into account such a special dust environment, we calculate the echo emission of FBOT and use it to model the near-IR excess discovered in the AT 2018cow event as well as its temporal evolution. The fittings of the spectra at different epochs are obtained with a unified parameter set. The spectral evolution of the echo emission can naturally be explained as a result of the evolution of the irradiating FBOT emission, instead of invoking some inexplicably varying parameter. Furthermore, in comparison with our results, the mass and distance of the dust shell derived with the traditional method could be usually underestimated significantly because the dust cooling and the temperature distribution have not been accurately reflected in the traditional fittings. In future nearby FBOT events, longer-duration IR observations will likely place more constraints on the properties of their dusty environment, since some more complicated processes would take place. For example, approximately a hundred days after an FBOT explosion, the ejecta will finally collide with and gradually engulf the whole dust shell, during which the dust grains could be partially destroyed and simultaneously experience recondensation.

Finally, by assuming a canonical gas-to-dust mass ratio of 100∶1 \citep{kochanek2011astrophysical,tartaglia2020long}, we can estimate that the total mass of the CSM of the AT 2018 cow is $\sim3\times10^{-2}\,M_\odot$, which can be much higher if the gas can be distributed at much smaller radii than the dust. Meanwhile, the duration of the mass-loss history of the progenitor could be approximately several hundreds to thousands of years, which is obtained by dividing the outer radius of the CSM of $\sim5\times10^{17}$ cm over the velocity of the progenitor wind. Here, the wind velocity could range from $\sim(20-30) ~\rm km~s^{-1}$ of red supergiants \citep{gonzalez2023effect},  $\sim300 ~\rm km~s^{-1}$ of luminous blue variables \citep{drissen2001physical} to $\sim(100-2500) ~\rm km~s^{-1}$ of Wolf$-$Rayet stars \citep{gal2014wolf}. Then, very roughly, the mass-loss rate of the progenitor can be constrained to be around $\sim(10^{-6}-10^{-4})\,M_\odot\ \mathrm{yr}^{-1}$, which is still very uncertain. These results somewhat point to a massive progenitor such as a Wolf$-$Rayet star. However, the low mass of the explosion ejecta ($\sim 0.03M_{\odot}$) requires that the progenitor star cannot be very heavy, which is at most an ultra-stripped helium star if it is not a compact star. This contradiction indicates that the progenitor system of AT 2018cow may experience a very complicated late-time evolution such as binary interaction, in which case the CSM environment could be produced by the companion star or by the common envelope of the binary \citep[e.g.,][]{Hu2023}. Finally, it is still worth mentioning that a lot of radio emission had been actually detected after AT 2018cow \citep{ho2019at2018cow,margutti2019embedded,nayana2021ugmrt}, which is very likely arising from the interaction of the explosion ejecta with the CSM \citep[e.g.,][]{chevalier2006circumstellar,harris2016against,petropoulou2016radio}. So, the modeling of this radio emission would provide an independent constraint on the CSM, which needs to be investigated carefully. 

\acknowledgements
We acknowledge Ning Jiang for sharing his observational results of AT 2018cow from NEOWISE with us. This work is supported by the National SKA Program of China (2020SKA0120300), the National Key R\&D Program of China (2021YFA0718500), the National Natural Science Foundation of China (grant Nos 12393811 and 12303047), Natural Science Foundation of Hubei Province (2023AFB321) and the China Manned Spaced Project (CMS-CSST-2021-A12).

\bibliography{2018cow}{}
\bibliographystyle{aasjournal}

\end{CJK*}
\end{document}